\documentclass[12pt,preprint]{aastex}

\shorttitle{Hard X-ray Variability and Jet Activity in R Aqr}
\shortauthors{Nichols et al.}

\begin{document}


\title{Discovery of Rapid Hard X-ray Variability and New Jet Activity in the 
Symbiotic Binary R Aqr}

\author{J. S. Nichols\altaffilmark{1},  
J. DePasquale\altaffilmark{1}, E. Kellogg\altaffilmark{1},
C. S. 
Anderson\altaffilmark{1}, J. Sokoloski\altaffilmark{1}}

\affil{Harvard-Smithsonian Center for Astrophysics, Cambridge, MA  02138 }
\and
\author{J. Pedelty\altaffilmark{2}}
\affil{NASA/GSFC, Greenbelt, MD 20771}
\email{jnichols@cfa.harvard.edu}






\begin{abstract}

Two $Chandra$ observations of the R Aqr symbiotic binary system taken 3.3 years
apart show dramatic changes in the X-ray morphology and spectral 
characteristics
in the inner 500 AU of this system.
The morphology of the soft X-ray emission has evolved from a 
nearly circular region centered
on the binary system to an hourglass shape that indicates the
formation of a new southwest jet.    
Synchrotron radiation
from the new jet in contemporaneous VLA radio spectra 
implies the physical conditions in the
early stages of jet development are different from
those in the more extended outer thermal jets known to exist for decades in this system.
The central binary source has two X-ray spectral components in each of the two epochs,
a soft component
and
a highly absorbed hard component characterized by T $\sim$ 10$^{8}$ K
if fit with a thermal plasma model. 
The spectrum hardened considerably between 2000.7 and 2004.0,
primarily due to increased flux above 5 keV, suggesting a change in
the accretion activity of the white dwarf
on a timescale of a few years or less.
Point-source Fe K emission is detected 
at the position of the central binary system in both observations. 
While the earlier observation shows evidence of only a single emission peak 
near Fe K$\alpha$ at 6.4 keV, the later
observation shows a more complex emission structure 
between 6 and 7 keV.  
Finally, we have discovered a modulation in the  hard
X-ray flux with a period of 1734 s at a 95\% confidence level
in the 2004 observation only.  The modulation potentially arises from 
standing shocks in an accretion column and we have explored the possibility
that the white dwarf
in R Aqr is analogous to the magnetic white dwarfs in Intermediate Polars.

\end{abstract}


\section{Introduction}

R Aqr is a symbiotic system composed of a Mira variable (M7III) and a white dwarf (WD).
  The  
Mira variable has a well-established pulsation 
period
of 387 days \citep{MA79}.  The orbital period has
been more difficult to determine, with periods of 44 yr 
\citep{WGM81} and 20 yr \citep{Kea89} often cited, although the 
period determination is quite uncertain
\citep{W86}.  The existence of a hot companion
to the red giant
was suggested in optical data \citep{M35,M50}, but confirmation
required the availability of the ultraviolet (UV) data from the International
Ultraviolet Explorer ($IUE$) \citep{Mea80,Kea86}.
Polarimetry of the central source \citep{Dea87} also confirmed the
binary nature of R Aqr. 
The binary system is associated with a compact H {\small II} region that 
is probably a dusty nebula produced by the Mira winds and ionized
by  the hot companion star UV radiation.
R Aqr is a D-type symbiotic system, defined as having comparatively red
infrared (IR) colors indicative of associated dust, as opposed to an S-type
symbiotic system with IR colors consistent with an isolated field red giant.
The temperature of
the WD in R Aqr is estimated at T=50,000--60,000$^\circ$ K, with a 
comparatively small radius of 3 $\times 10^{-3}$ R$_\sun$ \citep{MK95}.
The distance is  $\sim$200 pc \citep{vL97}.

R Aqr is remarkable because of its jet-like structures 
that extend more than 1400 AU from the central source. 
A jet in the R Aqr system was first discovered in 1977 in optical
wavelengths, appearing as a series of bright emission knots
about 6\arcsec\ NE of the central object \citep{WG80}.  A radio counterpart to 
the jet was found by \citet{Sea82}.  \citet{Hea91} derived a kinematic age of
about 90 yr for the NE jet.  A symmetrical jet in the SW direction
about 10\arcsec\ from the central object was detected by \citet{Kea89};
this jet is also suggested to be about 90 years old \citep{Hea91}.
A stream of emission knots extends from both jets to the central object,
implying collimated jet production occurring on several occasions in the
last 90 yr.  In the inner region within about 1\arcsec\
of the central binary, 
evidence of the initiation of jet ejections toward the
NE and SW has been seen in radio data \citep{Kea89,Dea95}.

 \citet{KM82} suggested that the jet production is associated with
 periastron of the system when the WD would be accreting
 matter from the wind of the Mira star at its maximum rate, resulting in the
 ejection of collimated high energy material from the accretion disk.
 In fact, \citet{PH94} found that the base of the northeast (NE) jet extends
 to within 15 AU of the Mira,
 confirming the central binary system as the source of the periodic
 jet production in this complex system.  These authors also found
 that the jet is highly collimated even very close to the binary system,
 with an opening angle of about 15$^\circ$.

 Collimated outflows, or jets, are associated with disk accretion onto
black holes in both binaries and active galactic nuclei, neutron
stars, and protostellar objects \citep{L97}.  Jets are also
seen in two of the three major classes of accreting WDs
(symbiotic stars and supersoft X-ray sources), as well as the hot
central stars of some (possibly binary) planetary nebulae \citep{Cowleyetal98,
Corradi01,Sokoletal04}.  Although
cataclysmic variable stars (CVs) produce bipolar disk winds, they have
not been found to generate well-collimated jets \citep{KniggeeLivio98,
Hillwigetal04}, perhaps because the accretion rate onto the WDs
in CVs falls below some minimum value needed for jet production (e.g.,
\citet{LasotaSoker05}).  Two symbiotic-star jets have so far been
detected in the X-rays: R Aqr \citep{Kea01} and CH Cyg \citep{GS04}.  
\citet{GS04} noted that the spectral energy
distributions of the CH Cyg and R Aqr outer jets are quite similar.
The inner regions of WD X-ray jets, however, have heretofore not been
explored.  The only WD X-ray jet extended sufficiently to resolve
the inner region is R Aqr.

 The high
resolution of $Chandra$ imagery allows us to analyze the morphological
and spectral changes of the R Aqr system on a small spatial scale in
the X-ray regime, isolating
changes in the jets from changes associated with the binary system.
Analysis of the outer jet structures in the R Aqr system as seen in the $Chandra$
data is reported in Kellogg et al. (2006, submitted).
 In this paper we focus on the $Chandra$ observations of the central binary
and the surrounding structure within 500 AU of the position of R Aqr.  
In Section 2, we present the $Chandra$ observations for two epochs, 
2000.7 and 2004.0.  
Section 3 describes the X-ray spectra of the central region and the new SW jet
as well as model fits to the central source spectra.  The analysis techniques used to 
identify the rapid variability of
the hard X-ray emission are described in Section 4.
Section 5 presents the 
VLA data
that were acquired contemporaneously with the 2004.0 $Chandra$ data.
Interpretations of 
the data are discussed in Section 6.
Conclusions are stated in Section 7.

\section{Observations and Data Processing}


An observation with the NASA $Chandra$ X-ray Observatory of R Aqr was obtained on 2003 December 31 (2004.0)
using the Advanced CCD Imaging Spectrometer (ACIS) with the target
placed on one of the 
back-illuminated CCD chips, designated S3.  
The dataset, obsid 4546, had an exposure time of 
36,534 s.  
In this paper, we compare this observation with an earlier 
observation 
of R Aqr taken 2000 September 10 (2000.7) that also used the ACIS S3 detector.  
The earlier 
observation, obsid 651, had an exposure time of 22,718 s \citep{Kea01}.  
Both observations were made with ACIS timed, very-faint mode and have a pixel size of 
0\farcs492.  In ACIS timed mode the CCDs are read with  a frame time of 3.2 seconds.
The very-faint mode evaluates 5 $\times$ 5 pixel regions 
centered on the pixel of interest to determine the actual count rate of each pixel, 
using a grade
system to determine the probability that a count is an actual X-ray event and not
an artifact. 
The data from both of the observations were further processed using the 
$Chandra$ Interactive Analysis of Observations (CIAO)
analysis tools to 
apply consistent calibrations and data processing techniques. The calibration 
files used in this data processing are those available with Calibration
Database (CALDB) 3.2.1. 
The event files were corrected for contamination of the optical/UV filter
on ACIS that has been increasing since launch, reducing the low-energy quantum 
efficiency,
and also corrected for a
time-dependent detector gain, which is a correction for a 
secular drift in photon energy
assignments.  The  Redistribution Matrix File (RMF) and Auxiliary Response File (ARF) 
were calculated specifically for each observation.

\begin{figure}
\plottwo{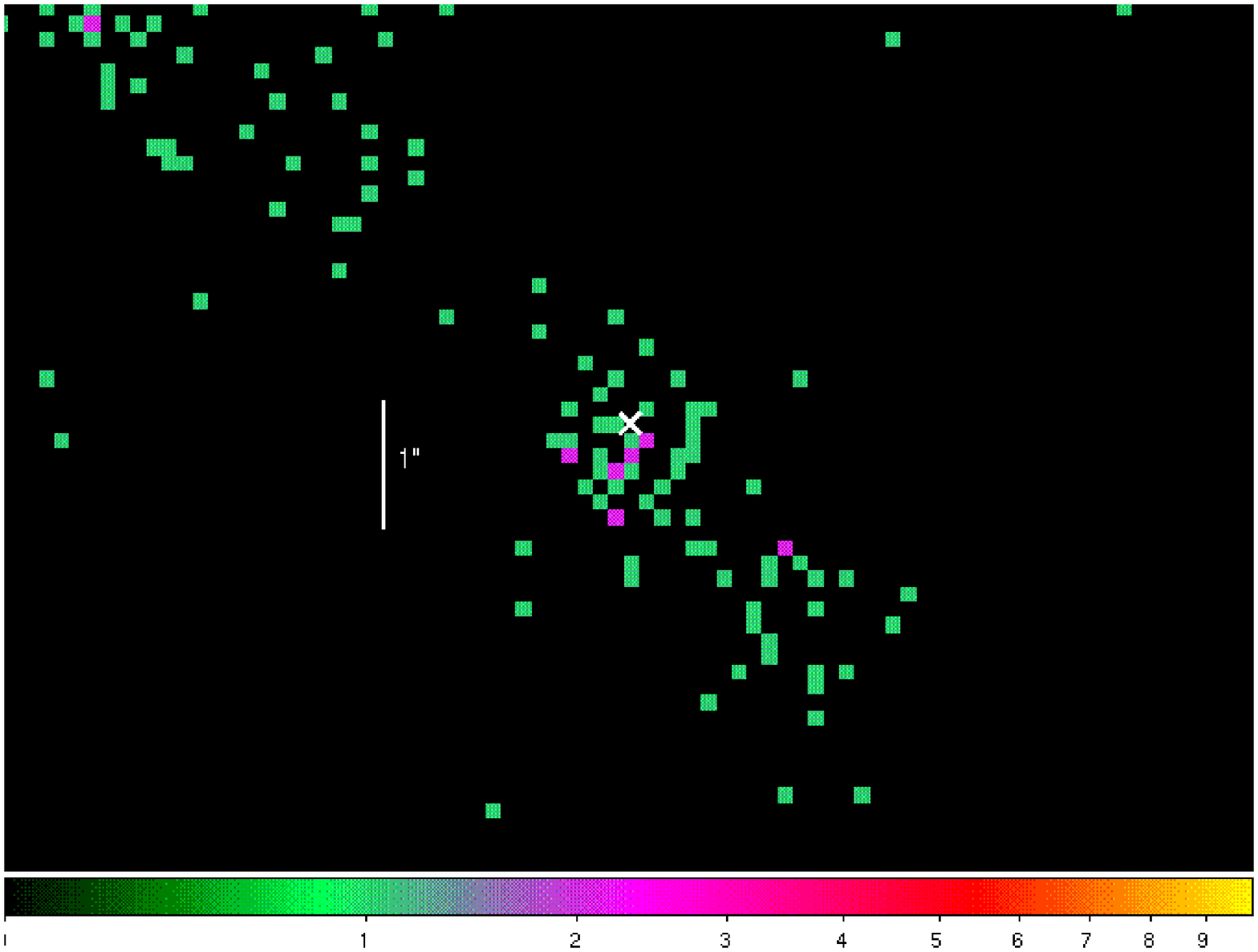}{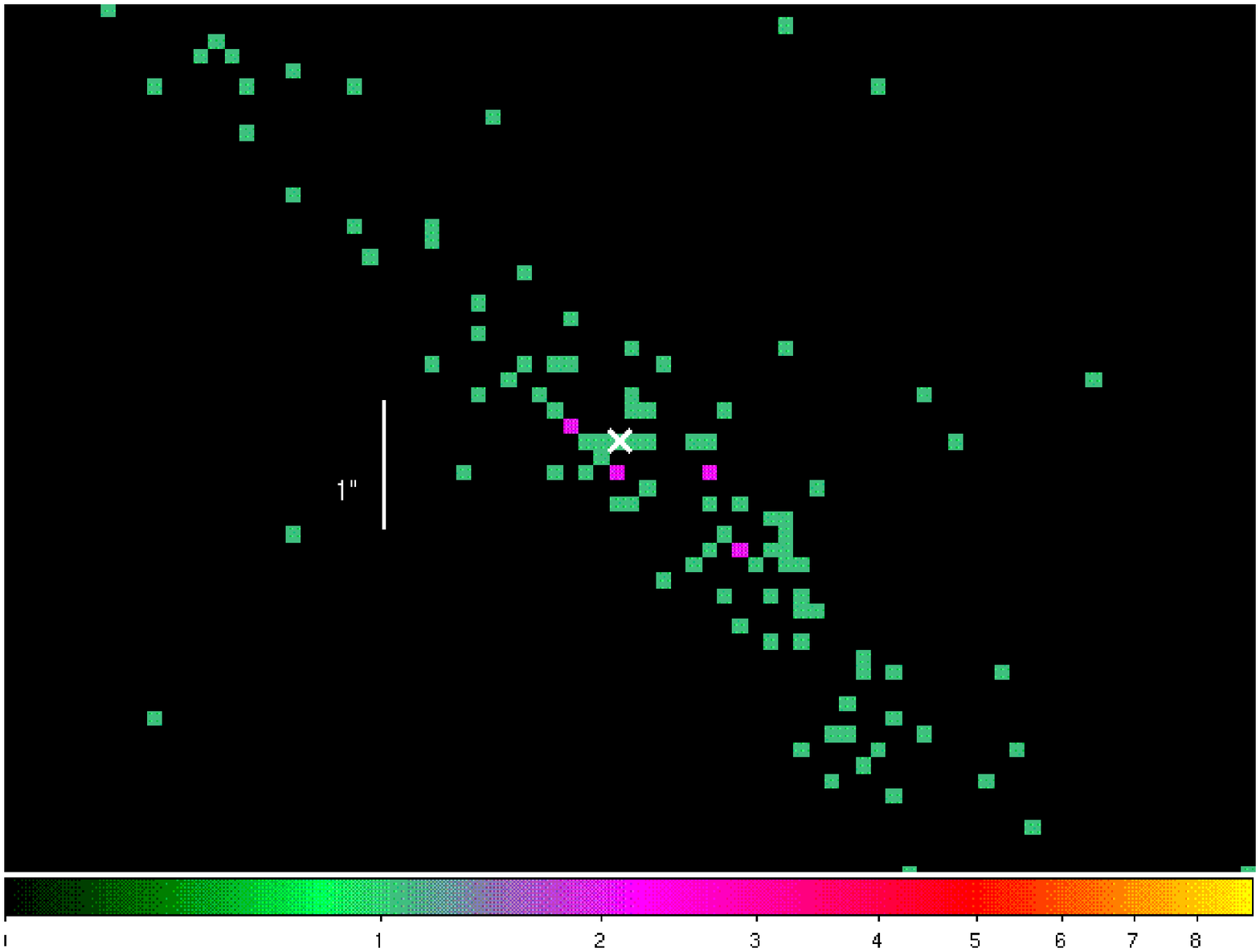}
\caption{Left:  Epoch 2000.7 image, filtered to display only events less than
2 keV.  The data were sub-sampled to 0.25 
pixels.  The position of R Aqr is marked with
a white "x", and the spatial scale of the image is indicated with a 1\arcsec\ bar,
which represents 200 AU at a distance of 200 pc. Right: Image of the 2004.0 data
prepared the same way as the 2000.7 image.  
The colorbar at the bottom of each image indicates the total
number of counts for each subsampled pixel.  
A new, extended soft X-ray component is apparent to the
southwest in the 2004.0 image.
\label{fig:sbs}}
\end{figure}

\begin{figure}
\plottwo{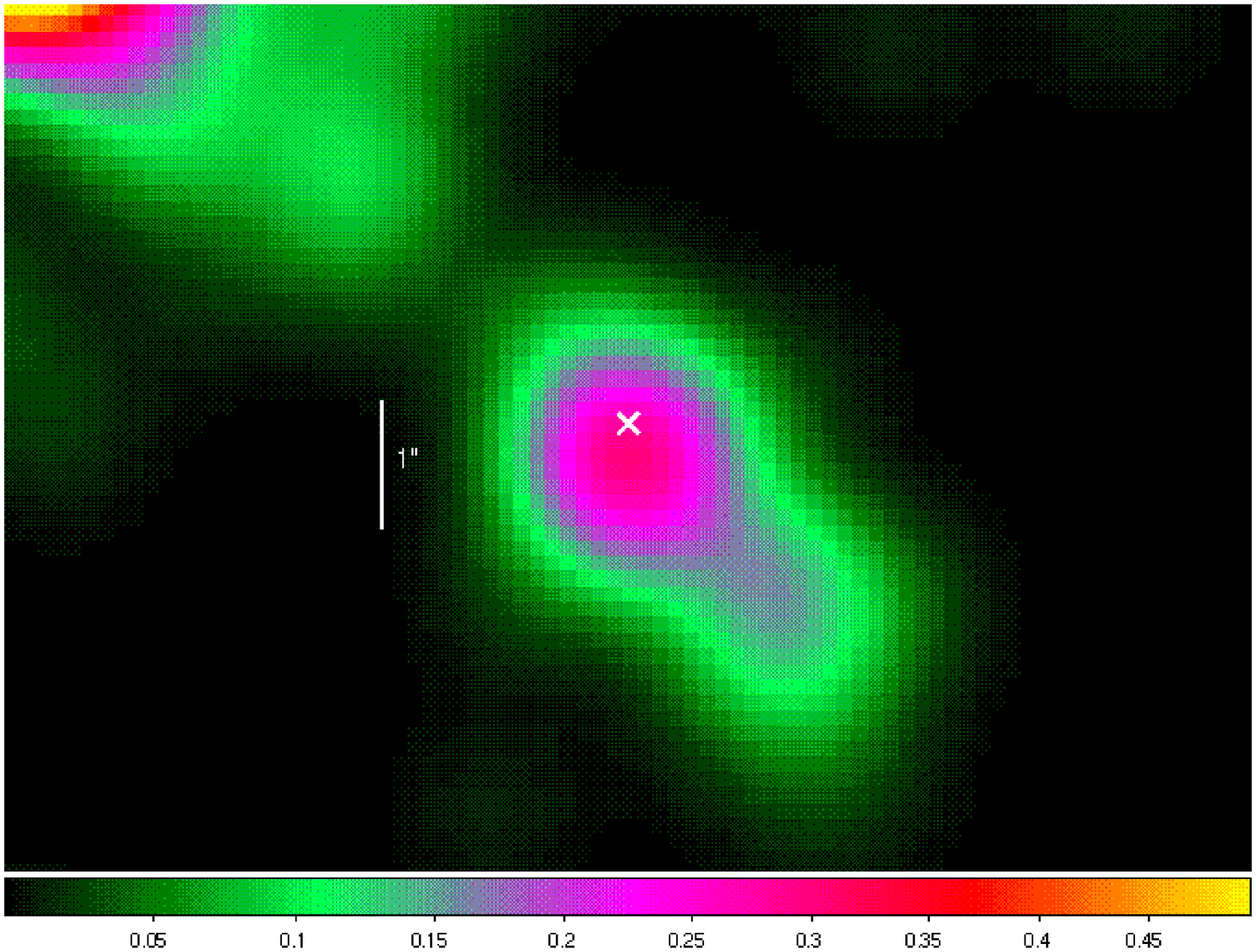}{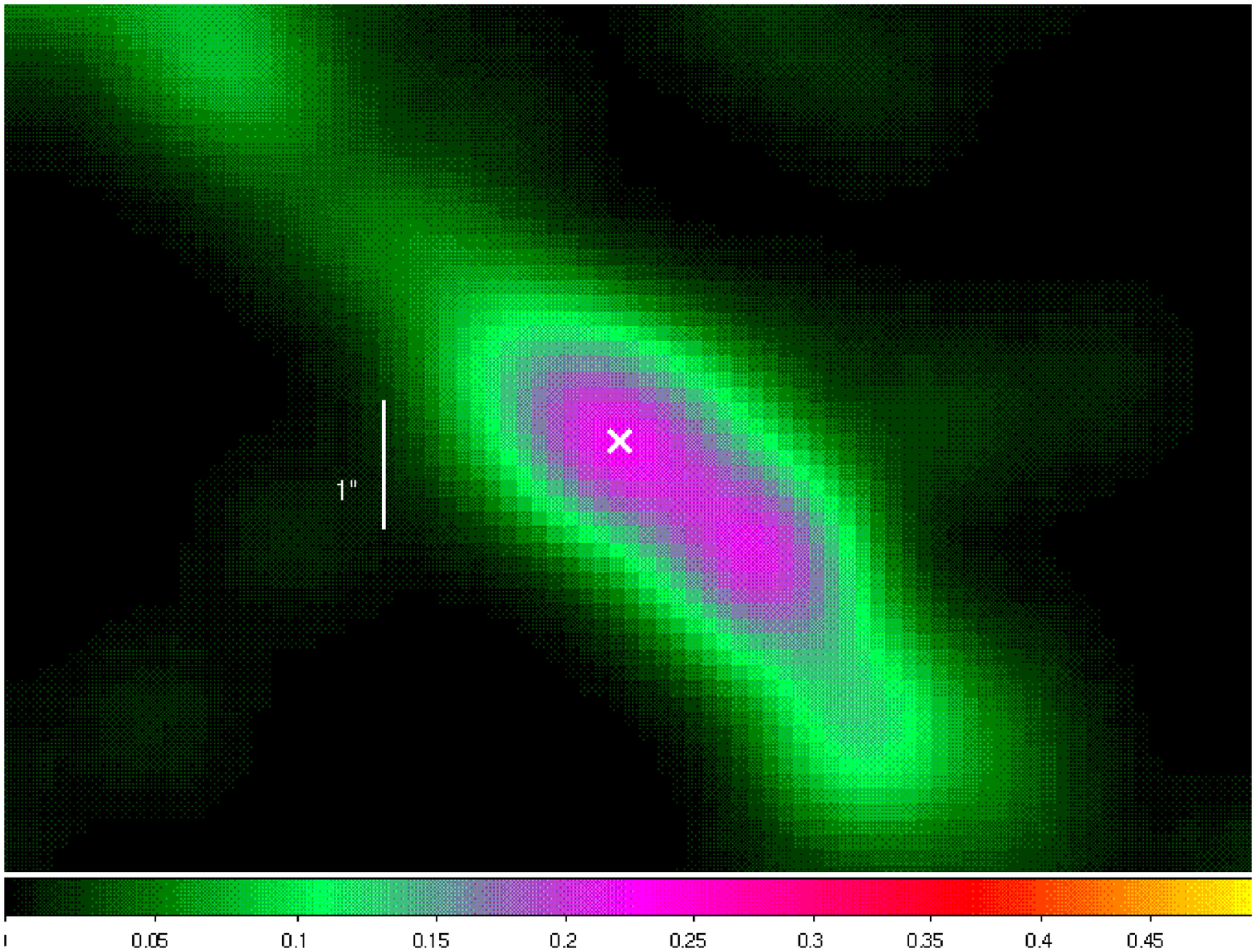}
\caption{Left: Epoch 2000.7 image, filtered to display only events less than
2 keV.  The data were sub-sampled to 0.25 
pixels and adaptively smoothed with a Gaussian kernel of radius 2 subpixels.  
The position of R Aqr is marked with
a white "x", and the scale of the image is indicated with a 1\arcsec\ bar, which represents
200 AU at a distance of 200 pc.  Right: Image of the 2004.0 data
prepared the same way as the 2000.7 data. 
The colorbar at the bottom of each image indicates the total
number of counts for each subsampled pixel. The new, extended soft X-ray component is 
more clearly apparent to the
southwest in this smoothed 2004.0 image.
\label{fig:sbssmooth}}
\end{figure}

Images of the soft X-ray 
(E $\le$ 2 keV) emission from the central region of R Aqr in 2000.7 and  
2004.0 
are shown in Figure \ref{fig:sbs}.  These images were created from the timed
event lists for each observation.  Using the floating point pixel position
determined by applying the time-dependent correction for the spacecraft
dither to the integer CCD chip pixel, sub-sampled images were created
such that each pixel in the subsampled image has dimension 0.25 $\times$
0.25 CCD chip pixel.  The subsampling is effectively a hardware subsampling,
with total counts preserved.
 The position of the Mira star (equivalent to the position of the
 binary system which is spatially unresolved at $Chandra$ resolution)
  is marked with an `x' 
on each 
image.  A correction to the position of the Mira star has been made 
based on proper motion in the Hipparcos database.  The accuracy of the 
$Chandra$ 
pointing is 1\arcsec or better, so the position of the binary system
is consistent with the position of
the soft X-ray peak flux.  Figure \ref{fig:sbssmooth}
presents adaptively smoothed versions of Figure \ref{fig:sbs}, created using
a Gaussian kernel of radius two subpixels with the CIAO csmooth tool.
The soft X-ray emission is clearly more 
extended in the southwest/northeast direction
in 2004.0 than in 2000.7.

\placefigure{fig:sbs}

\placefigure{fig:sbssmooth}

\clearpage


\clearpage

\begin{figure}
\plottwo{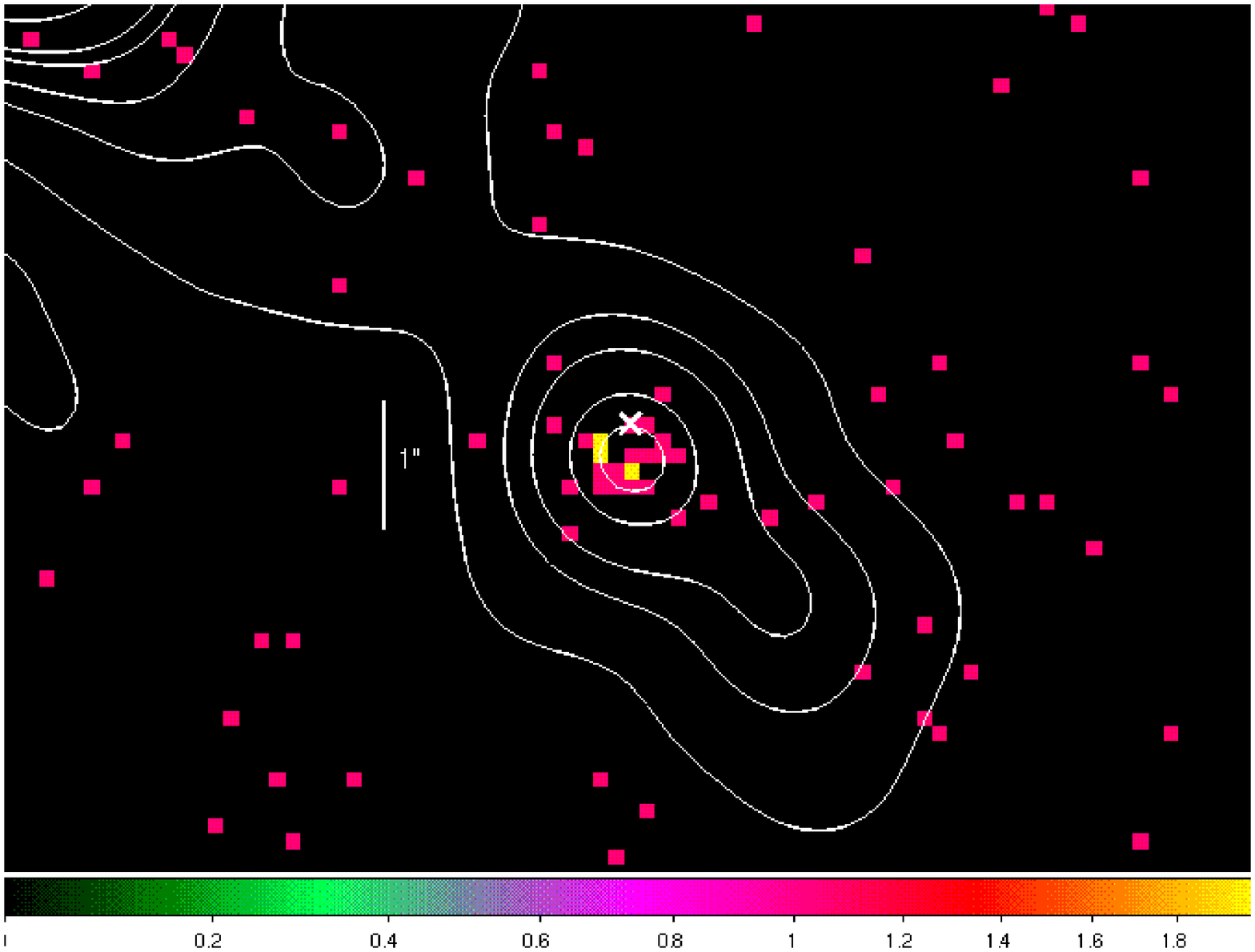}{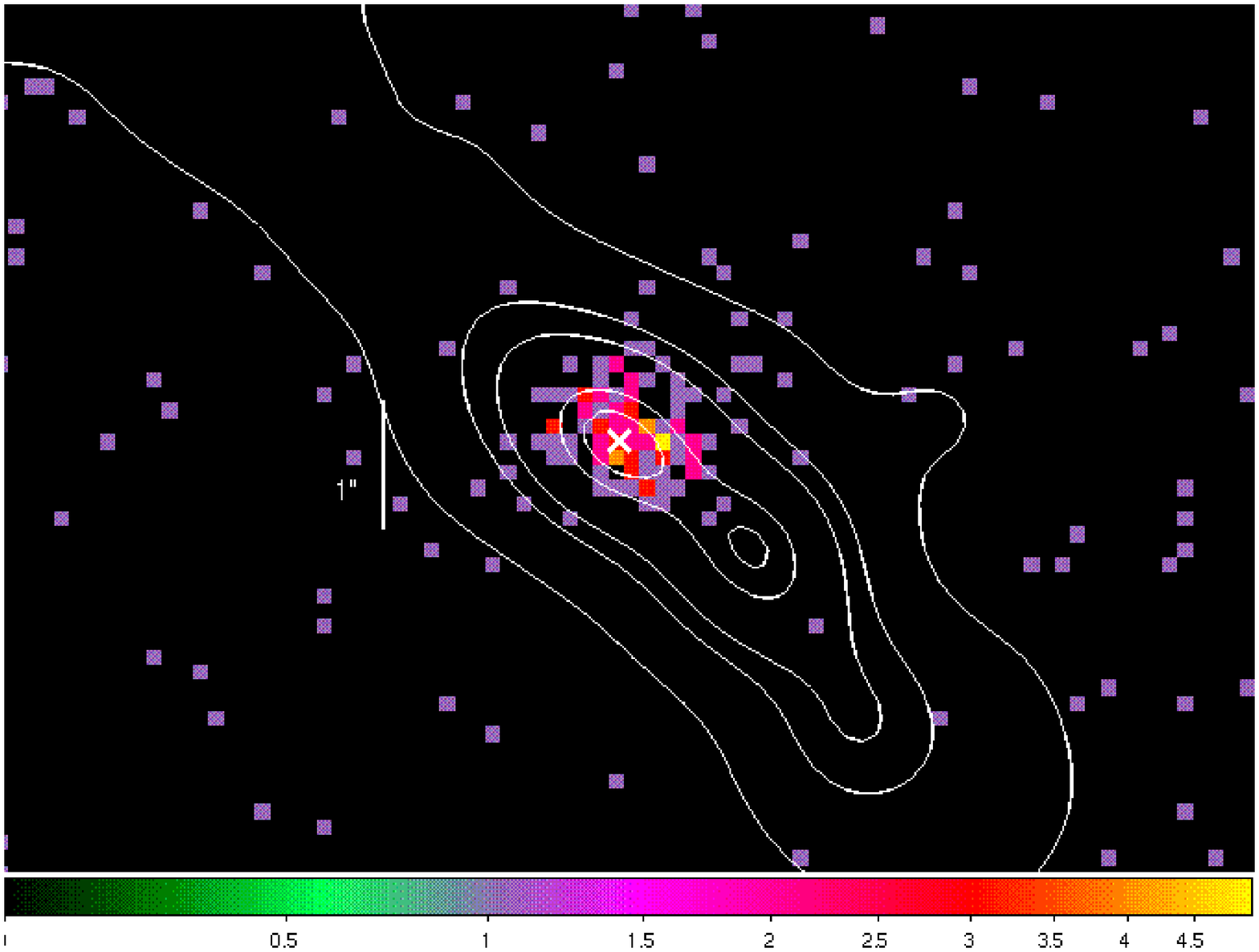}
\caption{Left: Epoch 2000.7 image, filtered to display only events greater
than 5 keV.  Contours from the smoothed soft X-ray
flux shown in Figure 2 are overplotted.  The scale is the same
as Figures 1 and 2, with 1\arcsec\ representing 200 AU at a distance
of 200 pc. The data were sub-sampled to 0.25
pixel, but were not smoothed.  The position of R Aqr, corrected for
proper motion, is marked with a white ``x''.  Right: Epoch 2004.0 image prepared the same
way as the 2000.7 image. The colorbar at the bottom of each image 
indicates the total
number of counts for each subsampled pixel. \label{fig:hard_651}}
\end{figure}

Images of hard X-ray emission
(E $\ge$ 5 keV) of the 
central region for both observations are shown in Figure \ref{fig:hard_651}
with contours of the smoothed soft X-ray
emission superposed for reference.   For this figure, we show only the 0.25 pixel subsampled 
images 
without adaptive smoothing, due to the small spatial extent of the hard 
X-ray 
region.  Although the spatial extent of the hard emission appears larger 
in 2004.0
than in 2000.7, in both cases it is consistent with 
a point source based on deconvolution of the point spread function with the measured
flux.  

\placefigure{fig:hard_651}

\clearpage
\clearpage

\section{X-ray Spectra}

\subsection{Central Source Spectra}

For each epoch, we extracted spectra from an elliptical
region at the position of the binary and another elliptical region
circumscribing
the extended soft emission to the SW.  These two regions
do not overlap spatially.  We first discuss the spectra extracted from
the region centered on R Aqr,
with x-axis = 2\farcs0, y-axis =
1\farcs3, and position angle  31$^{\circ}$ south of west. 
The total number of counts in the spectra for the central source is low for both the 
2000.7 observation (83 counts) and 2004.0 (195 counts).  There are too few counts
in either spectrum to bin the data at 10 counts per bin or higher in order to use
$\chi^2$ statistics in evaluating the spectral fits.  A bin
size of 10 counts or higher leaves
too few bins to yield a reasonable representation of the spectral morphology.
Based on the significant changes between the 2000.7 and 2004.0
spectra in counts per second as well as in hard X-ray
flux compared to soft X-ray flux, summing the spectra from the two epochs
together to increase the statistics is not considered justified. 

We used the C-statistic with the individual spectra binned at 3 counts per bin to
evaluate the spectral fits.  The background spectrum was fit independently
from the source flux.  Then the background components, scaled
in normalization by the ratio of the background area to the source area,
 were fit simultaneously with the source flux.  Simultaneous fitting of the background
and the source is appropriate for C-statistic analysis rather than
background subtraction because the C-statistic technique assumes Poisson
statistic errors in the counts, which would be violated with background-
subtracted data \citep{A96}. The background flux 
 is approximately
a factor of 200 less than the source flux, so the influence on the fit
is very small. 

The 2004.0 central source spectrum with 195 counts was used to evaluate 
several
spectral models to fit the data because the 2000.7 spectrum has too few counts
to justify use in the selection of a model fit.  For each trial fit,
the goodness-of-fit calculation for the C-statistic as found in XSPEC was run with 2000
iterations, so that the percent of realizations parameter is expected
to be accurate to about 2\%.  The soft portion of the 2004.0 central
source spectrum (0.25-2.5 keV) was fit with three simple models including 
interstellar absorption: (1) powerlaw;
(2) thermal emission APEC model; and (3) blackbody emission.  The thermal plasma 
model consists of an emission
spectrum from collisionally ionized gas, calculated using the APEC
code v1.10 as implemented in XSPEC.   All three of these
models gave an acceptable value of best-fit C-statistics and percent of 
realizations using the C-statistic in XSPEC.

The hard portion of the 2004.0 spectrum (2.5-10 keV) was first fit with four models including
interstellar
absorption: (1) powerlaw; (2) thermal emission APEC; 
(3) blackbody emission; and
(4) powerlaw plus thermal emission APEC.  We rejected all of these four models
for the hard emission
using the rejection criterion defined as more than 90\% of realizations 
less than best-fit C-statistics  (Arnaud, K., 2006,
private communication).  These four models each had 100\% of
realizations less than the best-fit C-statistics of 158.34, 371.92, 
199.74, and 158.71, respectively, resulting in rejection at greater
than 2 sigma.
However, five other models gave acceptable fits
for the hard portion of the 2004.0 central source spectrum,
all of which included interstellar absorption and a Gaussian Fe K$\alpha$ line:
(1) ionized absorption $\times$ thermal emission APEC;
(2) powerlaw; (3) neutral absorption $\times$ powerlaw; (4) neutral 
absorption $\times$ thermal emission APEC; and (5) neutral absorption
$\times$ blockbody emission.  The first of these acceptable models has 
6 free parameters, 
models
(3)-(5) have 5 free parameters each, and model (2) has 4 free parameters. 
We note that model (2) has a powerlaw component with a negative slope.

In summary, we find that the central source spectrum for the 2004.0 data does not have sufficient
counts to discriminate between several simple models.  However, we can eliminate
models that do not include a Gaussian component for the Fe K$\alpha$ line.
Also, all but one acceptable model for the hard component 
included additional absorption, either neutral or ionized.  
The exception was the powerlaw plus Gaussian model that
 had the largest C-statistic
of all the acceptable models
and a negative slope of -2.5.
We selected the interstellar absorption $\times$ powerlaw to fit the soft portion of
the spectrum.
Not only did the interstellar absorption $\times$ powerlaw have the lowest value
for the percent of realizations, but the discovery of radio synchrotron emission 
supports the selection of this model (Section 5).  The interstellar absorption
$\times$ (neutral absorber $\times$ (APEC thermal emission plus Gaussian for Fe K$\alpha$))
was chosen for the hard portion of the spectrum because it had the lowest value
of percent of realizations and best-fit C-statistics.  However, any of the acceptable models could have been used.  We present
this fit as a reasonable possibility for the spectral fit.

\begin{figure}
\includegraphics[angle=270,scale=.36]{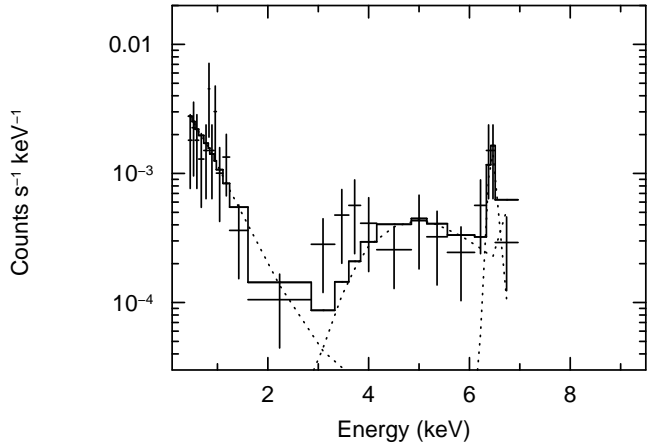}
\caption{Spectrum fit of central source of R Aqr in 2000.7.
The y axis is counts uncorrected
for instrumental response.   
Data were binned at 3 counts per bin.  
The fit consists of
interstellar absorption $\times$ a power law for energies less
that 2.5 keV.  For energies greater than 2.5 keV the fit is interstellar 
absorption $\times$ (a neutral absorber $\times$
a thermal plasma model (apec) with a Gaussian
profile to represent the Fe K$\alpha$ line).   The model components 
are shown as dotted lines.  Parameters of the fit are
listed in Table 1.
\label{fig:651center}}
\end{figure}

\begin{figure}
\includegraphics[angle=270,scale=.36]{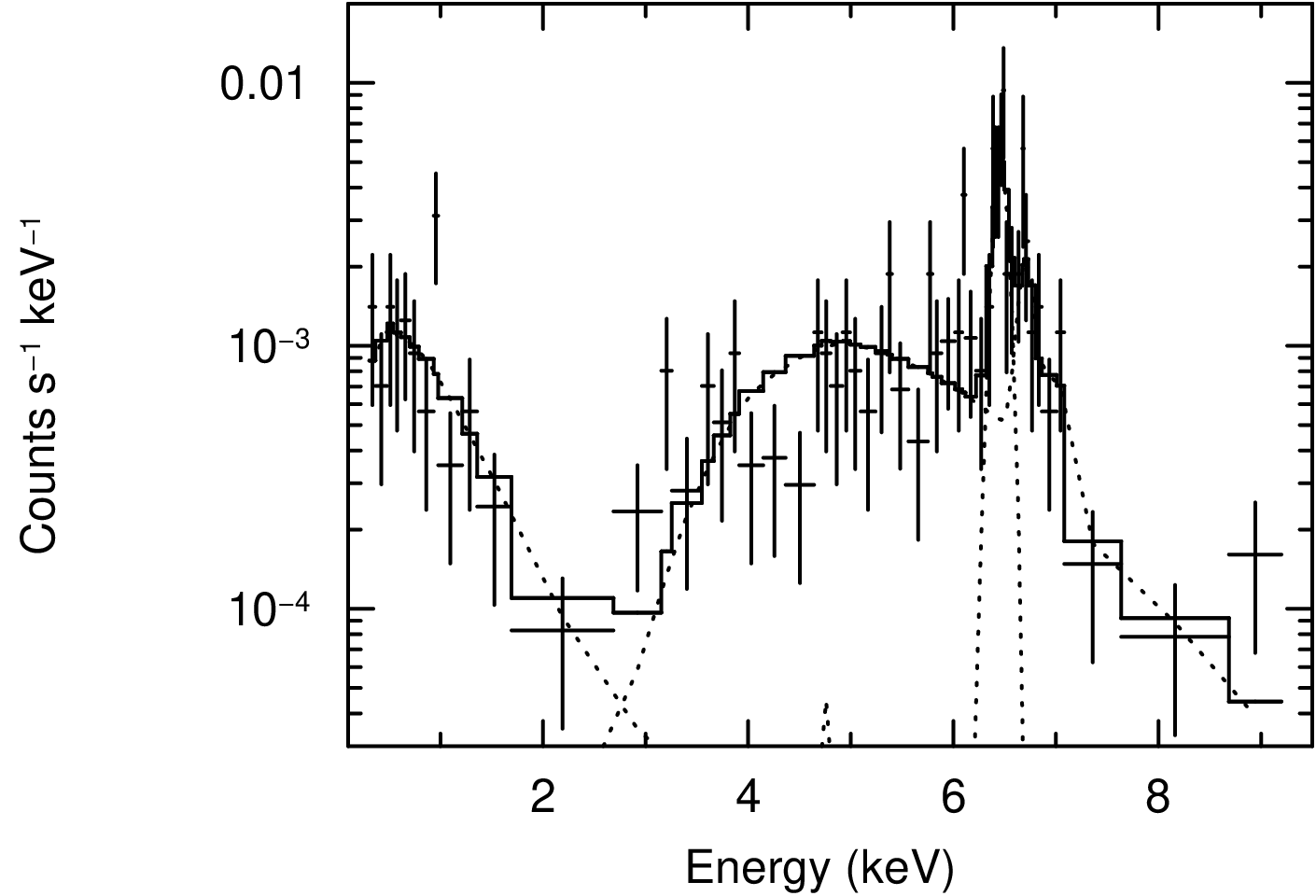}
\caption{Spectrum fit of central source of R Aqr in 2004.0.
The y axis is counts uncorrected
for instrumental response.   
Data were binned at 3 counts per bin.  
The fit consists of
interstellar absorption $\times$ a power law for energies less
that 2.5 keV.  For energies greater than 2.5 keV, the fit is interstellar 
absorption $\times$ (a neutral absorber $\times$
a thermal plasma model (apec) with a Gaussian
profile to represent the Fe K$\alpha$ line).  The model components 
are shown as dotted lines.  Parameters of the fit are
listed in Table 1.   \label{fig:4546center}}
\end{figure}

Because the 2000.7 data have fewer counts and thus weaker statistics for a fit determination,
we  fit the 2004.0 observation first, then froze several of the fit 
parameters from the 2004.0 
fit  to constrain the fit to the 2000.7 spectrum. 
Frozen parameters in the 2000.7 data were within acceptable ranges of those
obtained when we allowed them
to vary.
In addition, we ignored the background in our fit to the 2000.7 data because we expect only 
0.43 background counts with the 83 total counts.
 The parameter values from the fits to
each observation are listed in Table 1.  The line-of-sight interstellar absorbing
column based on 21 cm observations to R Aqr is 
1.85 $\times$ 10$^{20}$ cm$^{-3}$ \citep{Sea92} 
and 
the interstellar absorption 
parameter for all fits was
frozen at this value. We experimented with allowing the absorption to vary, but
always got an upper limit consistent with this column density.  Integrating these models, we
find $L_x$ (0.3-8 keV) of the central binary of $\sim$ 4 $\times$
10$^{29}$  $\times (200/d)^2$ erg s$^{-1}$ and $\sim$ 8 $\times$ 10$^{29}$ 
$\times (200/d)^2$ erg s$^{-1}$
in 2000.7 and 2004.0, respectively,
where $d$ is the distance to R Aqr in pc. 
The 2000.7 spectrum binned at 3 counts per bin and the fit determined as described above
are shown in Figure
\ref{fig:651center}.  The 2004.0 spectrum, also binned at 3 counts
per bin, and the fit are shown in Figure
\ref{fig:4546center}. 

\placefigure{fig:651center}
\placefigure{fig:4546center}

A thermal plasma  at the temperature determined from the 2004.0 observation, 
T=6.8 keV, includes
emission from He-like Fe {\small XXV} near 6.7 keV and H-like Fe {\small XXVI} near 6.97 keV.  
The He-like Fe {\small XXV} line is
suggested by visual examination of the  2004.0  spectrum, but there is no 
indication of the weaker H-like {\small Fe XXVI}
line. In the 2000.7 spectrum, 
the Fe {\small XXV} and {\small Fe XXVI} lines are included in the
thermal plasma fit by definition, but are not visually evident in the spectral data.  
In this spectrum, there are essentially no
counts with energy greater than 6.5 keV, so no conclusion can be drawn concerning 
either the Fe {\small XXV}
or Fe {\small XXVI}  
lines.  The Fe K  line detected in
the 2000.7 observation 
of the central core of R Aqr binary system
is consistent with emission from neutral Fe K$\alpha$, as discussed in
\citet{Kea01}.  
In contrast, the 2004.0 observation shows a more complex profile 
that might include both neutral Fe K$\alpha$ line and He-like Fe {\small XXV}. 
We emphasize, however, that there are too few counts in either the 2000.7
spectrum or the 2004.0 spectrum to reach meaningful conclusions
concerning validity of any particular model.
In both the 2000.7 and 2004.0
observations, the hard source is consistent with a point source, 
indicating this very hard emission
is within 100 AU of the binary system.

\subsection{Southwest Extension Spectra}

\begin{figure}
\includegraphics[angle=270,scale=.36]{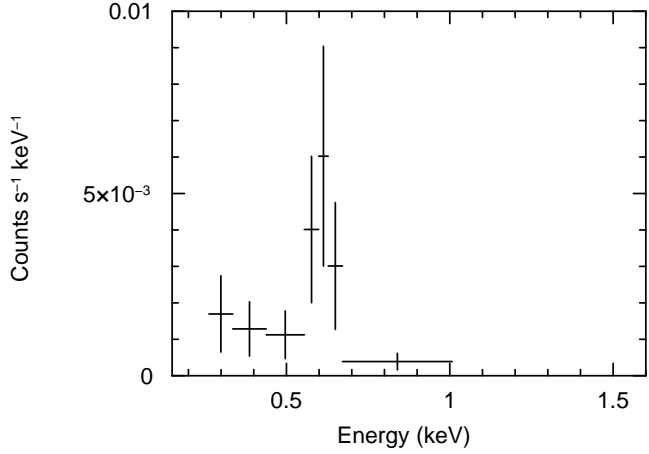}
\caption{Spectrum of SW extension region in 2000.7.  
The data are binned at 3 counts
per bin.
The y axis is counts uncorrected
for instrumental response.  There are no counts above 2 keV.
\label{fig:651sw}}
\end{figure}

\begin{figure}
\includegraphics[angle=270,scale=.36]{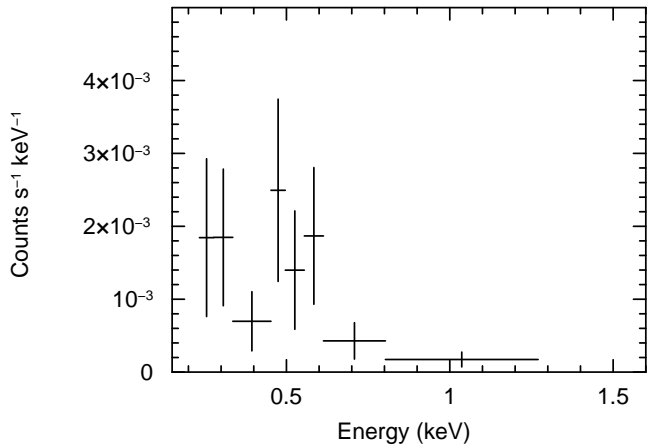}
\caption{Spectrum of SW extension region in 2004.0.  
The data are binned at 3 counts per bin.
The y axis is counts uncorrected
for instrumental response.  There are no counts above 2 keV.
\label{fig:4546sw}}
\end{figure}

The spectra of the SW extension were extracted for both epochs
with an elliptical region centered 
at RA = 23:43:49.4 and Dec = -15:17:06.1, with x-axis = 1\farcs8, y-axis = 
1\farcs4, and position angle of 31$^{\circ}$ south of west. 
The position of the extraction region was
centered on the peak of the soft X-ray emission of the SW extension in the 2004.0 observation.  
There is no flux at energies higher than 2 keV in either epoch.
The extracted spectrum for the 2000.7 data has only 21 counts
(see Fig. \ref{fig:651sw}) and thus cannot be fit with any confidence. 
Nor can the 2004.0 extracted spectral data with 29 counts 
be fit with any confidence.
However, an excess around the energy of O {\small VII} (0.57 keV)
is seen in the earlier epoch but absent in the 2004.0 inner jet spectrum 
(Fig. \ref{fig:4546sw}).  We cannot exclude a powerlaw fit to either
of the inner jet spectra.  Although no formal fit has been attempted, we
made the following observations from examining the spectra.  In 2000.7 the spectrum
of the SW extension is dominated by apparent emission near 0.57 keV, the energy
of O {\small VII}.  In 2004.0, there is no increase in flux over the entire
range compared with 2000.7, but the 0.57 keV emission is much weaker 
and possibly not
present.

\clearpage

\placefigure{fig:651sw}
\placetable{Table 4}
\placefigure{fig:4546sw}
\placetable{Table 5}

\clearpage

\section{X-ray Timing Analysis}

\begin{figure}
\includegraphics[scale=0.5]{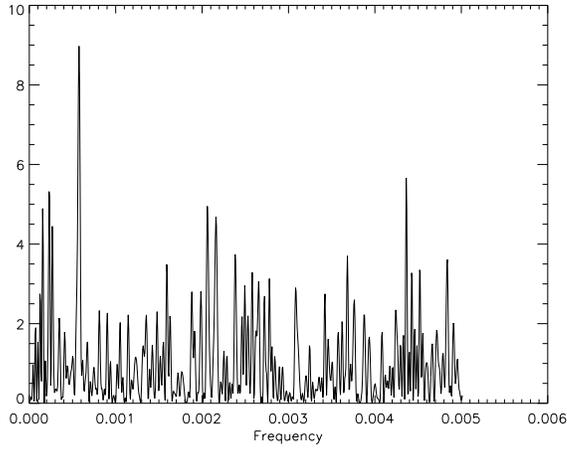}
\caption{Lomb normalized periodogram of the R Aqr binary source event data in the
2004.0 observation. Only photons with energies above 5.0 keV were considered.  The
period detected is 1734~s with a 95\% confidence level.\label{fig:frequency}}
\end{figure}

Rapid variability of the X-ray flux in energies greater than 5 keV was 
detected in the
2004.0 dataset for the region immediately surrounding the binary.
The event times were first corrected to barycenter times.  
The data were then binned at 100 seconds
per bin for the timing analysis and background-corrected.  The errors in the count rates were 
calculated using Poisson statistics 
for each bin of 100 seconds and represent one standard deviation.
We neglected any uncertainty in the photon arrival times 
because the 100 s bin size is significantly larger than the 3.2~s
frame time.
Using the Lomb normalized periodogram algorithm \citep{P92}, we found a period of 
1734~s with 0.05 significance 
level for the probability that the period is spurious,
and a maximum Lomb power of 8.96, which takes into consideration the number of bins used. 
The resulting periodogram is shown in Fig. \ref{fig:frequency}.
A similar analysis of the 2000.7 data did not reveal a period of credible
confidence.

\begin{figure}
\includegraphics[scale=0.5]{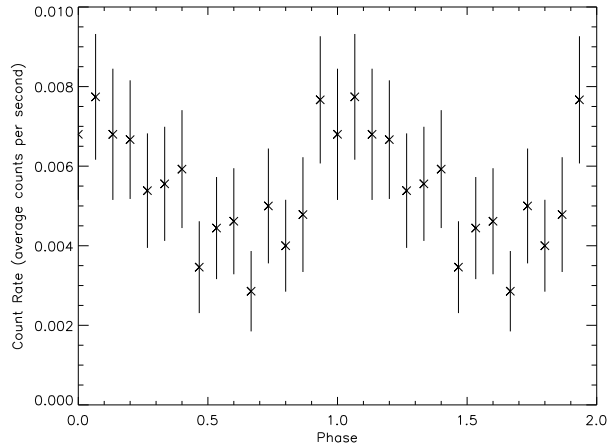}
\caption{Lightcurve folded on period of 1734 s ($\sim$29 m). 
Error bars represent one sigma errors and are based on Poisson statistics for 
the number of events in 
each 100 second bin, then
 propagated through
the folding process.
Only photons with energies about 5.0 keV were considered.  
Two phases have been plotted for clarity. Phase=0 is arbitrary.
\label{fig:lc}}
\end{figure}

As an additional check, we then performed a period search using epoch folding
for an array of trial periods
between 600 and 6000~s.
The period found with this search that gave the maximum $\chi^2$
result (maximum deviation implying the most likely period)
was 3475~s, approximately double the period determined with the Lomb 
periodogram.  This result could mean the period is double-peaked.
We accept the period determined by the Lomb periodogram of 1734~s
but with the caveat that the period could be 3475 s and double-peaked.
The lightcurve data 
for the 2004.0 observation were
folded on a period of 1734~s and
errors were propagated through the folding procedure.
The resulting folded light curve with associated errors 
is shown in Fig. \ref{fig:lc}, with two full phases shown in the plot
for clarity.  

\placefigure{fig:frequency}
\placefigure{fig:lc}

\clearpage

\clearpage

We evaluated whether the rapid variability could be related to the
dither pattern of the telescope during the observation.
The position of the central source
of R Aqr on the 2004.0 observation fell near a CCD node boundary
on ACIS.  The CCD node
boundary columns are less sensitive than other columns on the CCD.
The dithering of the telescope has a period of 1000 seconds
in the x-direction of the CCD chip and 707 seconds in the y-direction 
of the CCD chip.
The central source occasionally fell near or on one of the 
less sensitive columns 
at the edge 
of the dither pattern.  To determine if dithering onto the node boundary
influenced or even caused the variability we detected, we calculated
the percentage of the source flux that was on the CCD chip, rather
than on the node boundary, as a function of time, thus determining the effect of
the dither pattern.  We found a period of 1000~s for the
source being less than 100\% on the CCD chip, with an amplitude
for that period ranging from 98\% to 80\% of the source flux during the
course of the observation.   This 1000~s period is due to the
dither pattern.  No other period was detected 
for the spatial
position of the source as a function of the dither pattern.  As a further check,
we then
exposure-corrected  the events so that the events that had been placed in 
the node boundary due
to the spacecraft dithering had the same flux as they would have
had if they had fallen directly on the CCD chip node.  
Using this exposure-corrected
event list, we repeated the timing analysis and determination of significance 
of the variability.
There was no difference in the results between this exposure-corrected
light curve and the original light curve.  We believe this analysis
eliminates any instrumental cause for the variability we have detected.

\section{Radio data}

\begin{figure}
\includegraphics[scale=.4]{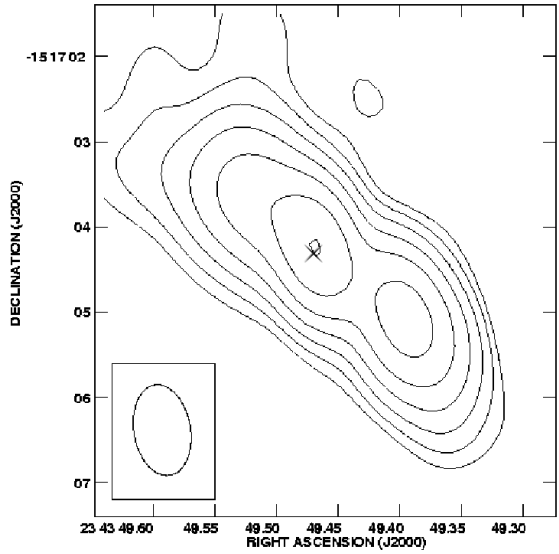}
\caption{3.5 cm image of the central source in R Aqr, acquired in
2004.0.  The peak continuum flux is 6.1 $\pm$0.2  mJy/beam.  The
contour levels are 1. $\times$ 10$^{-4}$ $\times$ 
(-2, -1, 1, 2, 4, 8, 16, 32 Jy/beam).  An ``x'' has been placed at the position
of R Aqr.
\label{fig:vlax}}
\end{figure}

\begin{figure}
\includegraphics[scale=.4]{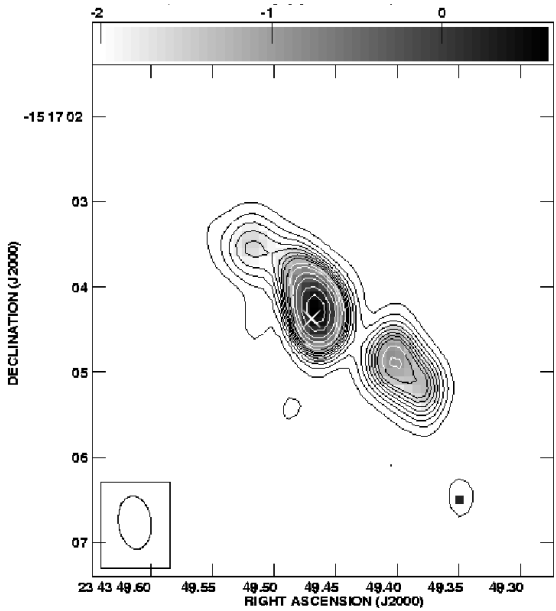}
\caption{2.0 cm contours of the central region of R Aqr, with peak
continuum of 8.61 $\pm$0.26 mJy/beam and contour levels of
2.0 $\times$ 10$^{-4}$ $\times$ (-2, -1, 1, 2, 3, 4, 5, 6, 7, 8, 10, 12, 
16, 20, 28, and 36 Jy/beam).  The contours are superposed on a gray-scale
image of the spectral index image. \label{fig:vlau}}
\end{figure}

\begin{figure}
\plottwo{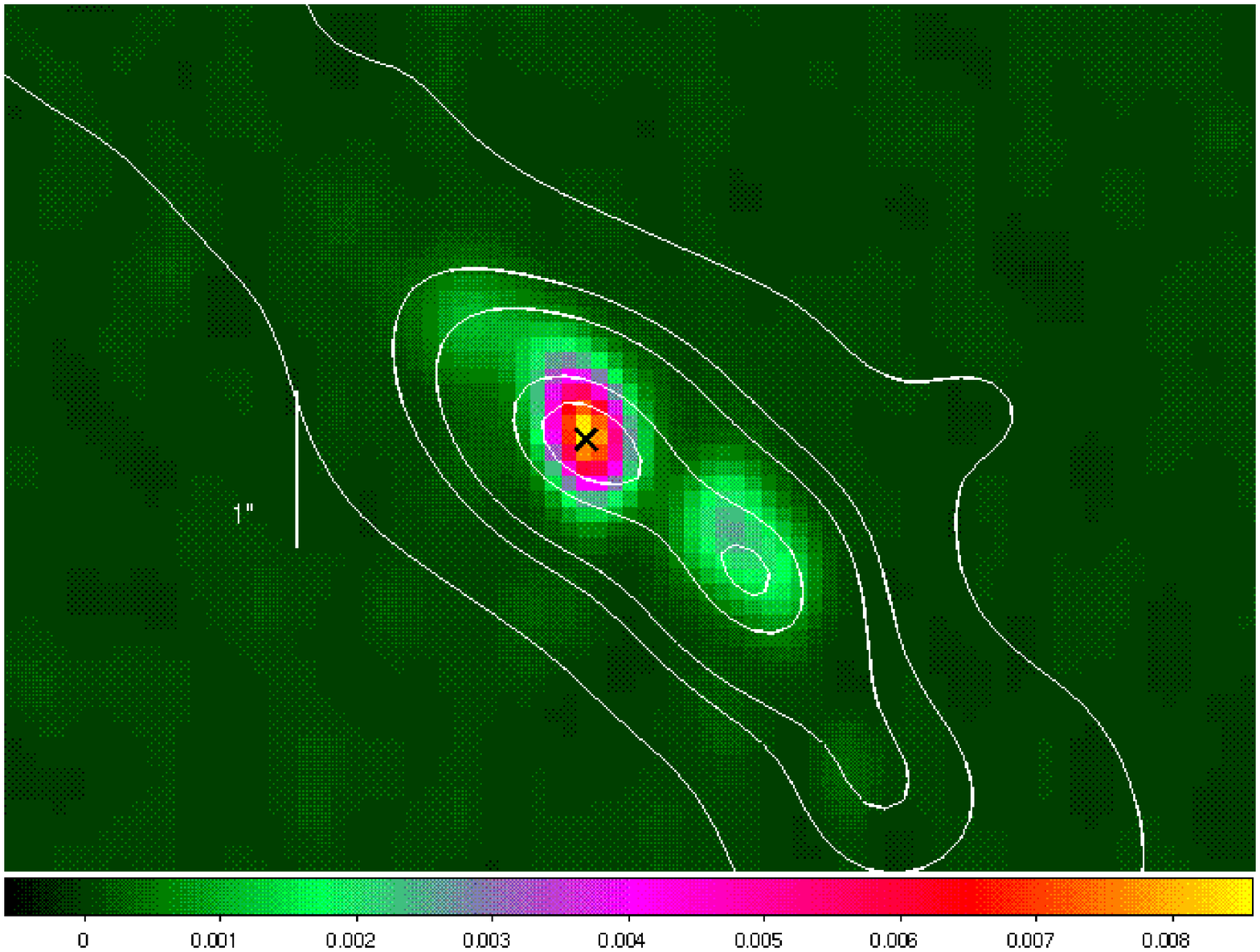}{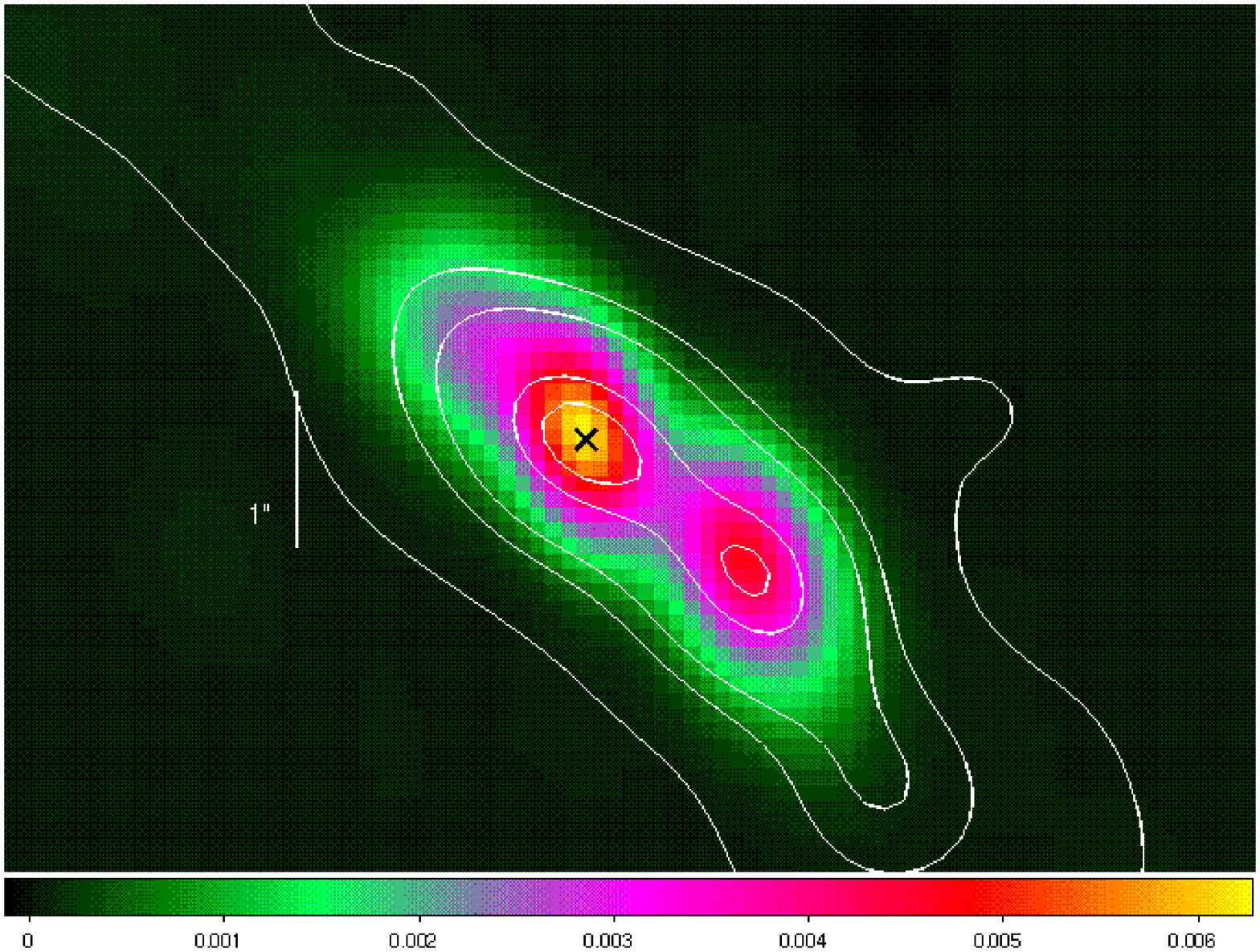}
\caption{VLA 2.0 cm (left) and 3.5 cm (right) observations with $Chandra$ 
2004.0 soft (X-ray $\le$2 keV) contours
superimposed.  The radio and X-ray emitting regions in the new SW extension 
are well aligned.\label{fig:vlax1}}
\end{figure}

R Aqr was observed on 2004 January 6 (2004.0) with the NRAO Very Large Array
in the B configuration.  
The receivers were tuned to the default U (2 cm) and
X (3.5 cm) frequencies with 100 MHz bandpasses.
Two scans were made at X band ($\sim$18 minutes each) and one scan 
at U band ($\sim$11 minutes).  We used the AIPS software for 
standard calibration, self-calibration, and imaging. Figure \ref{fig:vlax} 
shows
the contours and beam width for the 3.5 cm band observation, which
has a resolution of 
1\farcs0 $\times$ 0\farcs69, comparable to the $Chandra$ resolution.
The 2 cm band contours and beam width are shown in Figure \ref{fig:vlau}. 
The 2 cm band has a higher resolution of 
0\farcs63 $\times$ 0\farcs38, showing the radio detection of  
the new SW extension in finer detail.  The 3.5 cm and 2 cm band images are shown in
Figure \ref{fig:vlax1} as false color images instead of contours, with the soft X-ray
contours from the 2004.0 $Chandra$ observation superposed for reference.  
The alignment of the regions of maximum
flux in the radio and in the X-ray data is nearly exact, considering the $\sim$ 
0\farcs5 - 1\arcsec\
 pointing accuracy in the $Chandra$ data.

\begin{figure}
\includegraphics[scale=.4]{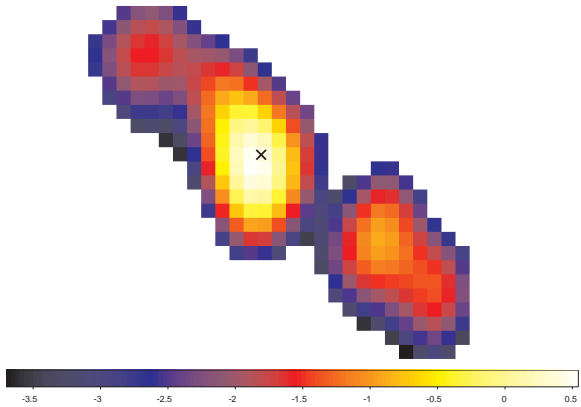}
\caption{Image of radio spectral index, determined from VLA 3.5 cm band and
2 cm band observations in 2004.0.    Negative values indicate synchrotron radiation.
The extremely negative values at the edges of the image are
edge effects of the algorithm. \label{fig:vla_sind}}
\end{figure}

Taking the radio spectral index, $\alpha_r$, to be defined such that
$F_\nu \propto \nu^{\alpha_r}$, where $F_\nu$ is the flux density at
frequency $\nu$, we determine the spectral index between 3.5 cm
and 2 cm from

\begin{equation}
\alpha_r = log(F_2/F_{3.5})/log(3.5/2)
\end{equation}
 
where $F_2$ is the flux density at 2 cm (15 GHz) in Jy and $F_{3.5}$ is
the flux density at 3.5 cm (8.6 GHz) in Jy. 
Using only data that allowed a determination of the flux densities
to greater than $5\sigma$, we constructed a map of the
spectral index (see Figure~\ref{fig:vla_sind}).
The 
spectral-index image allows the identification of regions of
synchrotron radiation, which have negative values of $\alpha_r$.  It
is clear from Figure~\ref{fig:vla_sind} that the central source is
thermal ($\alpha_r$ is positive) but the radio emission coinciding
with the new SW extension is non-thermal, as is a smaller region
NE of the central source.  This spectral-index image is also shown in 
grey scale
in Figure
\ref{fig:vlau}, superposed on the 3.5 cm contours.
Note that no soft X-ray emission is seen NE of the
central source in either of the Chandra observations.

\placefigure{fig:vlax}
\placefigure{fig:vlau}

\clearpage

\clearpage

\section{Discussion}

\subsection{X-ray Continuum from the Central Binary Source}

The X-ray count rate from the central binary increased between
2000.7 and 2004.0, due in an increase in the count rate of the
hard spectral component.   There was virtually no change in 
the count rate below 2.5 keV.  In terms of counts, we calculate
a hardness ratio of H/S where H = number of counts with energy greater 
than 2.5 keV
and S = number of counts with energy less than 2.5 keV.  Using this
definition, H/S $\sim$ 1 for  the 2000.7 central source spectrum
and H/S $\sim$ 4 for the 2004.0 central source spectrum.
The candidate model we explored in Section 3.1 indicates
the
thermal-plasma-fit flux increased by a factor of 2-3
(see Table 1) above 5 keV.  Below 2.5 keV, the flux either decreased or possibly
showed no change at the limits of the 90\% confidence measurements
for this candidate model.

The soft X-ray continuum for R Aqr
is likely  due
to an accretion activity or jet production.  
The soft X-rays may be due to reprocessed hard X-rays originating
on the surface of the WD if there is direct accretion onto the
magnetic poles \citep{Iea94}.
Another source of the soft continuum might be 
shocks generated by emerging jets \citep{GS04}.  
The soft X-ray ($\le$ 2.0 keV) could also be
attributable to an accretion disk.  
The hard continuum in the range 5-10 keV
can be produced from an accretion column if material from the
companion wind or an accretion disk is accreted directly onto the surface of the 
WD along magnetic field lines \citep{Cea85,Rea88}.
If the WD has
a sufficiently strong magnetic field, the accretion disk will not extend to the surface
of the star.  Instead, material flows from the accretion disk or directly from
the wind along field
lines to the magnetic poles of the WD, producing a standing shock
that can emit the hard X-ray continuum and emission lines. 
Another possible source of the hard X-ray continuum is an accretion
disk corona.  Material dissipating from the surface of an accretion disk
can form a hot corona around the WD star which produces
a hard X-ray continuum.
The hard X-ray continuum might also be due to 
colliding winds.  If the winds of the two stars have sufficient momentum, 
the 
collision of the winds can produce a bow-shaped shock and a wind-wind
interface that 
heats the gas to high X-ray
temperatures.  \citet{KT05} have modeled the colliding winds in
symbiotic stars and find that some systems will have  parameters consistent 
with
a colliding wind scenario.  Here we
explore the possibility that the hard X-ray emission from the central
binary in R Aqr is related to magnetic accretion. Both the
possible presence of $T\sim 10^8$~K gas indicated by the apec model and
potential for the He-like Fe line, as well as
the modulation of the hard X-rays at a period of 29~m, provide
hints that the WD in R Aqr might be magnetic.

The central source spectra are highly absorbed, with 
a n$_H$=$\sim3.5 \times 10^{23}$ cm$^{-3}$
determined from the candidate fit we chose for the 2004.0 spectrum. 
All of the acceptable models we tried required large amounts of
absorption, except the powerlaw + Gaussian which had a negative
slope which might represent a crude absorption model. Such a large amount
of absorption would be consistent with the line of sight passing
through pre-shock gas in the accretion column and/or an accretion curtain.  
Intermediate Polars (IPs) generally have heavily-absorbed
X-ray spectra with n$_H$ $\ge$ 1 $\times 10^{22}$ cm$^{-3}$, with
the notable exceptions of EX~Hya, V1025Cen, and HT~Cam \citep{Eea05}.
\citet{DM98} have discussed absorption in CVs in terms of multiple
absorbers from the pre-shock gas.  An accretion curtain absorber is
more distant from the WD surface and creates a more uniform absorption
of the shock emission region.  The dense wind of the Mira star can
also contribute large amounts of absorption.  We cannot differentiate
between these possibilities with the Chandra data presented here.

\subsection{Fe K emission}
                                                                                                                                                                                                                                                                                                                                                                                                                                                                                                 
The candidate spectral model that we have fit to the data 
of thermal emission for the hard X-rays includes He-like and
H-like Fe emission.
The presence of He-like Fe {\small{XXV}}
would indicate that the emitting region could have a temperature of
$T \sim (3-10) \times 10^7$~K, which is consistent
with the thermal plasma temperature of 
$\sim 10^8$~K from our model fit.  However, several other models are
equally viable.  The He-like Fe {\small{XXV}} line
is observed in the IP EX Hya. \citet{FI97}  attributed 
this thermal line to the standing shock
in the accretion columns above the magnetic poles .

Neutral Fe~K$\alpha$ lines are present in the spectra of the central binary source
in both epochs.  In the case of an accretion column, neutral
and near neutral iron fluorescence is expected from reflection of the
X-ray flux in the column on the surface of the WD \citep{DM98}, as
well as possibly in the post-shock and pre-shock regions of the accretion
column.
Another possible origin for the 
Fe K$\alpha$ neutral and lower excitation emission is the dense H~{\small II} 
nebula surrounding the binary system, formed
from mass loss by the Mira and ionized by the WD.
Given the known presence of 
jets in this system, the same collimated high energy material expelled by 
the WD/Mira system could cause fluorescence of Fe 
K as it impacts cooler, denser material while leaving the system.

\subsection{Rapid Variability of the Hard X-ray Emission}

Rapid variability of the hard X-ray emission  of 1734 s was detected 
with a 95\% confidence in the 
2004.0 observation for 
events with energies higher than 5.0 keV.  There are too few counts 
in the dataset to determine if the
variability detected is due to the hard continuum, the emission lines,
or both.   
The hard X-ray period of 1734~s would almost certainly represent the 
rotational 
period of the WD, and is similar to the WD rotation
periods in CVs and the magnetic symbiotic Z And
\citep[28~m;][]{SB99}.
If the WD in R Aqr does indeed have a strong enough magnetic field to
channel the accretion flow and produce the observed oscillation, the
WD spin period could be either 29~m, if the pulse profile is
single-peaked, or 58~m, if the pulse profile is double-peaked.  A
double-peaked pulse profile might indicate varying viewing angles of
an accretion curtain.
The rapid variability, coupled with the spectral morphology
similarities, indicates a
possible correlation between the WD in R Aqr and Intermediate Polars.
The variability is also strong evidence for a magnetic field, which 
is required for accretion onto a magnetic pole.  The rapid X-ray variability
is a compelling reason to reject a hot corona or a colliding wind  as possible
sources for the hard X-ray flux in R Aqr.

The currently accepted orbital period of 20-44~yr is too long to allow classification
as a CV.  A 20-44~year period  with
a rotational period of the hot companion of about 29~m would necessitate
a weak magnetic coupling of the
stars, possibly because of a large separation between the two stars.  While 
there are no mCVs with orbital periods as long as years, there are magnetic
symbiotic stars with such long orbital periods, such as Z~And 
\citep{Brocksopp04}.

\begin{figure}
\includegraphics[scale=.3,angle=90]{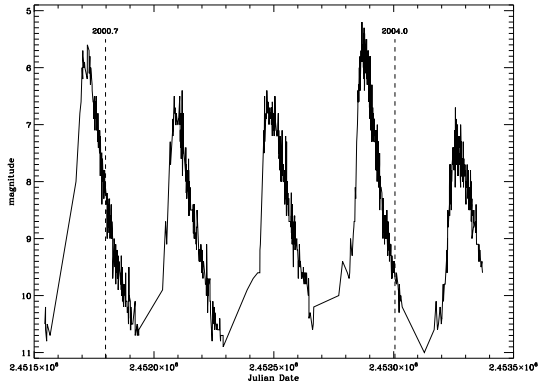}
\caption{Optical light curve from AAVSO with Julian dates of the 2000.7 and 2004.0 $Chandra$ observations
indicated.\label{fig:julian}}
\end{figure}

Although we have only analyzed two observations of 
R Aqr with $Chandra$ at this time, the first observation in 2000.7, taken about halfway between
optical maximum and minimum of the Mira star (see Figure \ref{fig:julian}), 
shows no detectable rapid X-ray variability and only a neutral Fe K$\alpha$
line, while the second observation in 2004.0, taken near optical 
minimum of the Mira star when mass loss would be at a maximum,
has $\sim$250\% increase in hard X-ray flux 
and rapid variability implying accretion onto the magnetic pole(s).  
A correlation between a change in the R Aqr WD accretion rate and/or mode,
and the pulsation
cycle of the Mira star is a tantalizing possibility that would require
much additional data at different phases of the Mira pulsation
period to explore.

\clearpage

\clearpage

\subsection{The New SW Jet}

It is apparent from Figures \ref{fig:sbs} and \ref{fig:sbssmooth} 
that a distinct emission extension in the SW has developed in
the soft ( $\le$ 2 keV) X-ray emission between the 
two observations.  
There is close agreement between the radio and the X-ray contours in
Figure \ref{fig:vlax1}.
The NE central binary soft X-ray emission is no longer circular but 
ellipsoidal in 
shape.  The SW extension has extended to about 125 AU from the position of 
R Aqr and 
is 
ellipsoidal in shape.  In the 2000.7 spectrum of the SW extension region, there is excess emission
in the 0.4-0.7 keV region of the spectrum, which could be O {\small VII} emission
at 0.57 keV associated with shock cooling.
Due to the change in morphology in 3.3 years 
from
a faint extension of the central region toward the SW to a
distinct elliptical region that rivals that of the central object 
itself in soft X-ray and radio
intensity, as well as the non-thermal radio 
spectrum in 2004.0, we interpret the SW region as the initiation
of a new synchrotron-powered jet.  

\placefigure{fig:vlax1}

The radio emission of the core of R Aqr has been monitored for more
than two decades \citep{KHM83,Dea95,Mea04}.   While earlier studies
find a clear extension to the NE of the binary position,  
\citet{Mea04}
find that the emission NE of the central source is decreasing
between 1995 and 1999.  A time series
of VLA images beginning in 1982, and obtained approximately
every four years, clearly shows the eruption of a new radio jet to the NE
in the late 1980's
(Pedelty, private communication).  
However, after 2000, 
the radio activity 
switched to the SW of the central object. The previous 
radio NE jet has 
faded considerably.
The peak intensity of the radio emission, 
centered on the star in each epoch, decreased by nearly a factor of
two in a little over three years.  In 2000 the central source was dominant, 
whereas in 2004 it is
more nearly equal to the intensity of the new SW jet.   In both 2000 and 2004
the radio spectrum of the central core is thermal, while the spectrum of the
SW jet is non-thermal in 2004.0.

\citet{Kea89} observed the inner region of R Aqr at 2 cm and 6 cm with VLA.
An $\approx$1\arcsec\ extension in the NE direction from the central
binary of R Aqr was found to be non-thermal in that study using data
taken in 1987.
Moving outward from the central object
along the projected trajectory of the NE jet, the radio emission 
quickly became thermal.
These authors speculated that the jet initiation mechanism 
is non-thermal, probably produced by
synchrotron emission from a mildly relativistic gas.  However, 
\citet{Dea95} report that in 1992-3 the radio flux of both the
small NE extension and a knot about 0.8\arcsec\ SW of the central source
were thermal and had increased in radio flux
by a factor of three in the $\approx$5~yr since \citet{Kea89} 
reported non-thermal
radiation.  Our detection of non-thermal radio emission in a new SW extension
is consistent with the theory that newly formed jets are non-thermal.
If the appearance of a new synchrotron-powered jet signals increased
accretion by the WD at periastron, then the data of \citet{Kea89} and this 
present study
suggest an orbital period of at most 17~yr.

Prior to the radio and X-ray observations reported here, \citet{MPK94}
predicted the formation of a new SW jet in the 2002 timeframe based
on redward-shifted UV emission lines with velocities of about +200 km s$^{-1}$
observed with $IUE$.
The appearance of these redward-shifted lines in 1992 was interpreted
as an indication of the initiation of a collimated wind or stream of
material which would begin to emit in radio, optical, and UV wavelengths about
10 years after the detection of the gas motion.  \citet{MPK94}
postulated that the jets in R Aqr are one-sided, alternating
direction on a timescale of $\sim$1 decade.  The new SW jet we have
identified appeared somewhere between 2000.7 and 2004.0, which
is right on the schedule \citet{MPK94} proposed.  Since only a jet
in the SW direction was detected in 2004.0, the \citet{MPK94} theory
of one-sided jet production is also supported by our observations.

Whereas most relativistic jets from neutron stars
or black holes produce non-thermal emission that in some cases extends
from radio to X-ray wavelengths \citep{FM04}, the
non-relativistic jets from WDs are primarily thermal X-ray emitters.
\citet{Kea01} found that the X-rays from the R Aqr outer jet
were consistent with thermal emission from a shock-heated plasma.
\citet{GS04} obtained a similar result for the CH Cyg
jet X-ray emission. The identification of non-thermal radiation in
the new, recently-formed SW jet makes it very unusual for a stellar
jet from a white dwarf and may yield important information about the formation and early evolution
of such jets.

\subsection{Comparison to Similar Objects}

The spectra of the central source of R Aqr presented here are similar to the X-ray spectra
of CVs, dwarf novae, and other symbiotic systems in that these objects
generally require a two-temperature component model and they commonly
show evidence of at least Fe K$\alpha$.  
Two-temperature components models such as we find for R Aqr 
are required 
to describe most of the ASCA spectra of mCVs \citep{ezuka99}, although
these objects commonly show more soft X-ray flux than R Aqr.     
The presence of Fe K lines, both fluorescent and thermal, in the spectrum is also common in CVs, magnetic CVs (mCVs), and 
symbiotic systems.
A survey of the ASCA data
of a sample of 23 mCVs showed Fe K$\alpha$ as well as thermal Fe lines in all of the spectra 
\citep{ezuka99}.

IPs are a subclass of mCVs in which the accretion disk is truncated at the
inner edge  and accretion takes
place at the magnetic poles, with the accreted matter transported from the disk
via accretion curtains.   Modulation in the X-rays due to the spin period is a signature of IPs,
as is a ratio of P$_{spin}$/P$_{orbit}$ $\ll$ 1.
The X-rays are produced primarily at the site of accretion in the accretion column and
WD surface rather than in the
boundary layer, which never reaches the surface of the WD.
Most IPs  have both a thermal 
and fluorescent contribution to the Fe lines \citep{Iea91}.  High levels of absorption are
generally required to fit the spectra of mCVs, including IPs.   The spectra of IPs have been fit with  plasma
models  and  partial or ionized absorbers (V709 Cas; \citet{dMea01}, BY Cam; \citet{Kea96}, 
TV Col; \citet{Rea04}).
The spin period of the WD is often detected in X-ray observations of IPs, but is
not always energy-dependent.  
TV Col, based on RXTE, ROSAT, and ASCA data \citep{Rea04}, does have an energy-dependent
spin period modulation, similar to what we find for R Aqr.  The orbital period can 
sometimes be derived from the
power spectra of IPs by detection of sidelobes and their harmonics.  We did not
find obvious evidence of sidelobes. IPs, as opposed
to polars, can have
spin periods that are much shorter than the orbital period because the 
magnetic field is weaker and not necessarily sufficient to lock the WD
spin to the orbital period.  The
X-ray data we have presented from R Aqr are similar to IP X-ray data due to the
possible presence of He-like Fe {\small{XXV}} and the detection of variability on the order of
less than an hour.   However, IPs have orbital periods on the order of hours,
not years.  Therefore, R Aqr is probably a magnetic symbiotic star analogous to,
but not identical to, an IP. An example of a magnetic symbiotic is Z And, 
which has a 28 min oscillation 
\citep{SB99} and a
small radio jet \citep{Bea04}

CH Cyg is a nearby symbiotic system that has several similarities to R Aqr.
CH Cyg has jets detected in the radio \citep{Tea86},
optical \citep{Crockerea01} and X-ray \citep{GS04}.  
While the donor stars in CH Cyg and R Aqr are of similar spectral type (CH Cyg:
M6III; R Aqr: M7III), the donor in R Aqr is a Mira variable.  
The $Chandra$ X-ray spectrum of the accreting central source in CH Cyg presented
by \citet{GS04} is quite similar
to that of R Aqr in 2000.7.  X-ray spectra of both stars required a two-temperature
fit to the continuum with a highly absorbing column.  A strong Fe K$\alpha$ line 
is present in the CH Cyg spectrum \citep{Eea98}.  \citet{GS04} find a shock-heated plasma at the interface
of the jet and the surrounding ISM.   \citet{Eea98} also concluded that 
the soft component of the continuum originates
from the shocks generated by the jets.   In the case of R Aqr,
we have modeled the X-ray spectrum for the central source separately from the
new jet, but shocks due to the emerging jets may contribute to the soft continuum
in the central source.      

Another interesting comparison can be made with Hen 3-1341, a symbiotic system with stellar jets.
\citet{Munari05} find that the jets develop during periods of increased accretion rate and 
cease to develop
as the wind from the 
WD decreases.   The implication is that jet production 
is coupled with accretion rate.  The jet production in 
R Aqr may also be 
dependent on the wind speed and mass loss from either the WD or the donor star.   
The spectral changes including
increase of the hardness ratio and the possibility of detection of He-like Fe 
{\small XXV} 
indicate a significant change
of state in the accretion process in R Aqr on a time scale of few years.

\section{Conclusions}

$Chandra$ X-ray observations of the symbiotic system R Aqr
have revealed unexpected variability on a time scale of minutes
as well as a time scale of a few years.  Identification of a 1734 s modulation
in the hard ($\ge$ 5 keV) X-ray emission 
in the 2004.0 observation 
with a confidence level of 95\% is a strong suggestion that the WD
in R Aqr has a magnetic field.
The rapid variability of the Fe lines and/or hard continuum, if verified by
future observations, implies an origin in the standing
shock above the accreting column at the magnetic poles of the WD
and is assumed to represent the spin period of the WD.

The spectral energy distribution and total flux also changed
over the 3.3 yr time period between the two observations.
We are not able to identify a unique acceptable model
due to the low number of counts in the two spectra analyzed.
We can, however, make some general statements based on the properties of
the spectra and the acceptable fits that we studied.  There is an increase 
in the hardness
ratio in the 2004.0 data compared to the 2000.7 data of a factor of 4.  
All acceptable fits for the hard portion of the spectrum require a
large amount of absorption, with the exception of a powerlaw with a negative
photon index.  Also, all acceptable fits required a Gaussian profile
to fit the Fe K emission.  
As an example, we selected a model with a
combination of a power law for the soft X-ray emission and a highly absorbed, 
collisionally excited, thermal plasma model
of temperature T = 10$^{8}$ K for the hard X-ray emission. Several other simple models
cannot be excluded considering the number of counts in the spectra. 
Based on our candidate model, the soft X-ray continuum has either decreased
in intensity or possibly not changed during the 3.3 yr interval between the 
$Chandra$ observations,
while the hard X-ray continuum has definitely increased. 

We explored the source of the X-ray emission for the case of the
candidate model we presented.
If the WD is indeed magnetic and the X-ray emission originates
in a standing shock at the magnetic pole(s), the soft X-ray component may be
due to reprocessed
hard X-rays at the surface of the WD, with some possible contribution
pre- and post-shock regions of the column and possibly from an accretion disk.
The hard X-ray continuum and emission lines would be produced by
the standing shock in this scenario.  The temperature we have determined for
the hard component of the 2004.0 spectrum is consistent with
an origin in a accretion column shock.
  For this scenario, we interpret the changes in the
X-ray emission as indications of changes in the accretion parameters
of the R Aqr binary.

During this 3.3 yr period, 
an extension of soft X-rays appeared that we conclude is evidence
of a new jet emitted in the SW direction from the central binary position. 
Contemporaneous radio data 
have non-thermal spectra indicative of synchrotron emission from the new jet.  
Previous radio observations of R Aqr in 1987 showed a non-thermal extension 
from the central object 
toward the NE, interpreted as a new jet.  Non-thermal radiation may be a 
signature of a newly emitted jet, even in stellar systems where older jets
are generally thermal.  The history of jet formation in the R Aqr
system suggests non-thermal emission during the earliest  part of the
jet evolution, and an orbital period of about 17 yr if the jet formation
is coupled with periastron.

The dramatic change in the X-ray emission properties and
the appearance of a new jet in the 3.3 yr time interval 
between the two $Chandra$ observations suggest a far more dynamic 
system than has previously
been assumed for R Aqr, and symbiotic stars in general.  
The data suggest the accretion mode and/or rate can change
on a timescale of a few years or less.  We find that R Aqr may have 
some similarities with Intermediate Polars
in the rapid variability and spectral characteristics.  
But the long orbital period of R Aqr indicates it is more likely a
magnetic symbiotic system.

\acknowledgments 

We are grateful for  constructive suggestions by the
anonymous referee and J. Grindlay.
We acknowledge the support of NASA grant GO4-3050A and
contract NAS8-39073.  The $Chandra$ X-ray Observatory is
operated for NASA by the Smithsonian Astrophysical Observatory.
The National Radio Astronomy Observatory is operated
by Associated Universities, Inc., under cooperative agreement with 
the National Science Foundation.

Facilities: \facility{CXO, VLA}.

\bibliographystyle{apj}
\bibliography{raqr}

\begin{deluxetable}{lll}
\tabletypesize{\footnotesize}
\tablecolumns{3}
\tablewidth{0pc}
\tablecaption{Spectral Fit Parameters: Central Source}
\tablehead{
\colhead{}    &  \multicolumn{2}{c}{Fit Parameters\tablenotemark{a,b}} \\
\colhead{Quantity} & \colhead{2000.7}   & \colhead{2004.0} }
\startdata
Total Counts &83& 195 \\
Absorption 1 (nH 10$^{22}$cm$^{-3}$)&0.0185 [frozen]&0.0185 [frozen] \\
Powerlaw Photon Index  &2.6 [frozen]&
2.6 [2.0,3.2] \\
Powerlaw Normalization (10$^{-6}$ ph cm$^{-2}$ s$^{-1}$)&1.9
[1.4, 2.4]&
1.4 [0.98, 1.8] \\
Absorption 2 (nH 10$^{22}$cm$^{-3}$)&35[frozen]&35 [28, 45] \\
Apec temperature (keV)&6.8 [frozen]&6.8 [4.2, 10.8] \\
Apec normalization (10$^{-6}$ ph cm$^{-2}$ s$^{-1}$) & 230. [155., 310.] & 
530. [360., 800.] \\
Fe K Gaussian line energy (keV)& 6.43 [frozen] & 6.43[6.41, 6.46] \\
Fe K Gaussian sigma (keV)& 0. [frozen] &0.0 [frozen] \\
Fe K Gaussian normalization (10$^{-6}$ ph cm$^{-2}$ s$^{-1}$ )& 3.0 
[1.1, 5.9] &9.6 [6.7,13.0] \\
C statistic &20.48 &118.94 \\
&(24 PHA bins,& (121 PHA bins, \\
&21 deg of freedom)&104 deg of freedom) \\
Goodness of fit & 21.25\% realizations&22.05\% realizations \\
&$<$ best fit stat 20.48&$<$ best fit stat 118.94 \\
Unabsorbed Source Flux (10$^{-15}$ erg cm$^{-2}$ s$^{-1}$) &88 [63,126]
& 210 [140,320] \\
~~~~~~~0.3--8 keV&& \\

\enddata
\tablenotetext{a}{C-statistic 90\% confidence level indicated}
\tablenotetext{b}{Solar abundances assumed}

\end{deluxetable}





\end{document}